\documentclass[twocolumn,showpacs,preprintnumbers,amsmath,amssymb]{revtex4}
\usepackage{graphicx}% Include figure files
\usepackage{dcolumn}% Align table columns on decimal point
\usepackage{bm}% bold math

\begin{document}

%\preprint{APS/123-QED}

\title{A Lower Bound on Neutrino Mass and Its Implication on The Z-burst Scenario}% Force line breaks with \\

\author{Kwang-Chang Lai}
 \email{kclai@phys.ntu.edu.tw}
\affiliation{Department of Physics, National Taiwan University,
Taipei, Taiwan.}

\author{Pisin Chen}%
 \email{chen@slac.stanford.edu}
\affiliation{Kavli Institute for Particle Astrophysics and
Cosmology,\\
Stanford Linear Accelerator Center, Stanford University, Stanford, CA 94309}%

%\author{W.-Y. Pauchy Hwang}
% \email{wyhwang@phys.ntu.edu.tw}
%\affiliation{Department of Physics, National Taiwan University,
%Taipei, Taiwan.}

\date{\today}% It is always \today, today,
             %  but any date may be explicitly specified

\begin{abstract}
We show that the cascade limit on ultra high energy cosmic
neutrino (UHEC$\nu$) flux imposes a lower bound on the neutrino
mass provided that super-GZK events of ultra high energy cosmic
rays (UHECRs) are produced from Z-bursts. Based on the data from
HiRes and AGASA, the obtained neutrino mass lower bound violates
its existing cosmological upper bound. We conclude that the
Z-burst cannot be the dominant source for the observed super-GZK
UHECR events. This is consistent with the recent ANITA-lite data.

\end{abstract}
\maketitle

\section{Introduction}

%Bounds on neutrino flux\cite{WB:1998yy} and AGN bounds\cite{MPR}

%GRB neutrino flux\cite{WB:1997ti}

Big bang cosmology predicts the existence of both cosmic microwave
background (CMB) and cosmic neutrino background (C$\nu$B). Ultra
high energy cosmic protons are expected to interact effectively
with the CMB photons, predominantly through the photopion
production at $\Delta$-hyperon resonance, and would lose their
energies rapidly with the attenuation length around $50$Mpc. As
such the ultra high energy cosmic ray (UHECR) spectrum is
predicted to exhibit a cutoff --- the so called GZK cutoff
\cite{GZK1,GZK2} --- around the threshold energy
$\sim4\times10^{19}\mathrm{eV}$. While observations from the HiRes
experiment is consistent with the notion of GZK cutoff
\cite{Abbasi:2002ta}, the AGASA data appears to suggest the
opposite \cite{Takeda:1998ps,Takeda:2002at}. This leads to many
speculations as to whether the GZK cutoff really exists, and if
not what is the nature of these super-GZK events.
\\

Existing models for super-GZK UHECRs are usually categorized into
top-down and bottom-up scenarios. The top-down scenario assumes
the existence of super-massive exotic elementary particles based
on theories beyond the standard model. The major challenge of this
scenario lies in the demand for a fine-tuned decay and/or
annihilation rate and the lack of physical evidence for their
existence. On the other hand, the bottom-up scenario, which
assumes ordinary particles as the UHECRs, faces the challenge of
providing an effective mechanism to accelerate particles to ultra
high energies. Even if an effective ``cosmic accelerator" can be
identified, the issue of cosmic transport dictated by the GZK
mechanism remains, as there does not seem to exist identifiable
sources within our local super cluster ($\sim 50$Mpc) for the
detected events.
 \\

To circumvent this difficulty, it was suggested that the Z-burst,
the resonant annihilation of the ultra high energy cosmic neutrino
(UHEC$\nu$) with the C$\nu$B into a Z boson and its subsequent
decay into ultra high energy protons
 \cite{Weiler:1982qy,Roulet:1992pz,Yoshida:1996ie}, that occurs
within our local super cluster can account for UHECRs beyond the
GZK-cutoff \cite{Fargion:1997ft,Weiler:1997sh}. With its
mean-free-path comparable to the present Hubble radius, the
UHEC$\nu$ serves as a cosmic messenger that can avoid the GZK
proton attenuation problem without invoking particle theory beyond
the standard model. For the Z-burst to happen, the UHEC$\nu$ must
be at a resonant energy,
\begin{equation}
E_{res}=\frac{M^2_Z}{2m_\nu}\cong
4\times10^{21}\Big(\frac{1\mathrm{eV}}{m_\nu}\Big)\mathrm{eV}\\,
\end{equation}
which depends on the neutrino rest mass, $m_\nu$. Here $M_Z$
denotes the mass of the Z boson. If the Z-burst mechanism is
indeed responsible for the observed UHECR super-GZK spectrum, then
there must exist a constraint on the neutrino mass via the above
relation.\\

Based on the Z-burst scenario, two groups have derived bounds on
neutrino masses using AGASA data with different strategies. Fodor,
Katz and Ringwald \cite{FKR:2001qy} deduce the Z-burst spectrum
from the AGASA data by parameterizing the transition from the
non-burst to the burst component near and above the ``ankle" of
the UHECR spectrum. Gelmini, Varieschi and Weiler
\cite{GVW:2004zb} derive their bound by requiring the Z-burst not
to overproduce non-observation events beyond the AGASA end-point
energy. Our strategy, instead, is to invoke an upper limit on the
UHEC$\nu$ flux so as to obtain an upper bound on the required
resonant energy $E_{res}$, which can in turn be translated into a
lower bound on the neutrino mass.\\

In this paper we derive a lower bound on the neutrino mass based
on the assumption that Z-burst mechanism saturates the observed
UHECR super-GZK spectrum. Our deduced neutrino mass lower bound,
however, turns out to be higher than the existing upper bound
deduced from cosmological considerations. We thus conclude that
the Z-burst mechanism cannot be responsible for the super-GZK
UHECR spectrum. Our conclusion agrees with that from the recent
ANITA-lite experiment \cite{ANITA}.

\section{UHECR Flux and Z-Burst}

%\begin{figure}[htb]
%\input{C:\Program Files\WinEdt Team\WinEdt\fig6.ps}
%\end{figure}

Assume that all observed super-GZK UHECR proton events are induced
from Z-bursts. The observed super-GZK proton flux must be smaller
than the total Z-burst proton yield in the universe since there
must be events occurred outside our local GZK-sphere ($\sim 50$
Mpc) which could not reach the Earth. Furthermore, in order for
Z-burst events to saturate the observed super-GZK spectrum, it is
inevitable that they are oversupplied since there must be some
Z-burst protons that are generated at energies below the
GZK-cutoff. Therefore,
\begin{equation}
I_{p,>GZK}^{obs}\leq I_{p|\nu}^{Z}\hspace{0.5mm},\label{main}
\end{equation}
where $I_{p,>GZK}^{obs}$ is the total observed super-GZK proton
flux with energy exceeding the GZK-cutoff and $I_{p|\nu}^Z$ the
total proton flux from Z-bursts, both in
units of $\mathrm{cm^{-2}s^{-1}sr^{-1}}$.\\

Though observations \cite{Shinozaki:2002ve} cannot completely rule
out the possible contribution to the super-GZK UHECR spectrum by
UHE photons, experimental data \cite{Shinozaki:2002ve} suggests
that protons saturate the super-GZK flux,
$I_{p,>GZK}^{obs}=I^{obs}_{>GZK}$, at $2\sigma$ confidence level.
Then in terms of the total observed UHECR flux, Eq.(\ref{main})
can be written as
\begin{equation}
I^{obs}_{>GZK}\leq I_{p|\nu}^{Z}\hspace{0.5mm}.
\end{equation}

%SuperGZK flux
The AGASA experiment has accumulated $57$ events above
$4\times10^{19}\mathrm{eV} $ with a total exposure of $\sim4
\times 10^{20}\mathrm{cm^2\hspace{0.5mm} s\hspace{0.5mm}
sr}$\cite{Hayashida:2000zr}. This translates into an observational
super-GZK flux,
\begin{equation}
I_{>GZK}^{obs}\simeq 1.43\times
10^{-19}\mathrm{cm^{-2}s^{-1}sr^{-1}}\hspace{0.5mm}.
\end{equation}
It can be shown that a fitting spectrum with a power-law index
$-2.78$ \cite{Nagano:2000ve} reproduces the above flux.\\

\section{Z-Burst Yield}

Now we deduce the total Z-burst proton yield within a relevant
cosmic volume. Solar and atmospheric data on neutrino oscillations
indicate that the oscillation lengths are much shorter than the
solar distance. So for cosmic neutrinos their population among the
$3$ flavors should be equalized. The total UHEC$\nu$ flux is thus
simply 3 times that for a single neutrino flavor. We further
assume that UHEC$\nu$ fluxes are the same for neutrinos and
antineutrinos. By definition,
\begin{eqnarray}
I_{p|\nu}^{Z}&=&3\xi_{p+n|\nu}
         \int_0^{R_{max}}dr\int_0^{\infty}dE F(E,r)\nonumber\\
&\times&\sigma_{\nu\bar{\nu}}(E=\frac{s}{2m_{\nu}})
\mathrm{Br}(Z\rightarrow\mathrm{hadrons})n_{\nu}(r)\hspace{0.5mm}.
\label{4}
\end{eqnarray}
Here $F(E,r)$ is the UHEC$\nu$ flux at energy $E$ and distance $r$
from the Earth, $n_{\nu}(r)$ is the number density of the C$\nu$B,
$\sigma_{\nu\bar{\nu}}(s)$ the neutrino-antineutrino cross section
at $s=2m_{\nu}E$, Br($Z\rightarrow \mathrm{hadrons}$) the
branching ratio, and $\xi_{p+n|\nu}$ the multiplicity of nucleons
per Z-burst.\\

For completeness, our integration should include all neutrinos and
Z-burst events in the universe. Such a treatment tends to be
over-conservative as the protons deduced from Z-bursts outside of
our local GZK-sphere may hardly survive. The complete but
ultra-conservative treatment is discussed in the appendix. A
physically reasonable yet much simplified calculation can be
carried out by neglecting the contributions outside of our local
GZK-sphere. This amounts to replacing the maximum distance
$R_{max}$ in our integration by the radius of our local GZK-sphere
$(R_{GZK}\sim 50{\rm Mpc})$. As the distance under consideration
is much more local, all the $r$-dependence can be ignored:

\begin{eqnarray}
I_{p|\nu}^{Z}&=&3\xi_{p+n|\nu}
         R_{GZK}n_\nu\mathrm{Br}(Z\rightarrow\mathrm{hadrons})\nonumber\\
         &\times&\int_0^{\infty}dE F(E)\sigma_{\nu\bar{\nu}}(E)
\hspace{0.5mm},
\end{eqnarray}
where $n_\nu(r)\equiv n_\nu=112{\rm cm^{-3}}$ is the
neutrino-antineutrino number
density per flavor at present.\\

%Simplification of energy integration
The UHEC$\nu$ flux is commonly assumed to follow a power-law
energy spectrum
\begin{equation}
F(E)=F_0 E^{-\alpha},
\end{equation}
with $F_0$ being the normalization factor.

Using $E=sE_{res}/M_Z^2$, we can write the energy integration as
\begin{eqnarray}
  &&\int_0^\infty dE F(E)\sigma_{\nu\bar{\nu}}(s=2mE)\nonumber\\
  &=&E_{res}\int_0^\infty
  \frac{ds}{M_Z^2}F(sE_{res}/M_Z^2)\sigma_{\nu\bar{\nu}}(s)\hspace{0.5mm}.\label{E:int}
\end{eqnarray}

As the neutrino-antineutrino annihilation cross section is sharply
peaked at the Z-resonance, it acts essentially like a
$\delta$-function in the integration over the energy of the
UHEC$\nu$. We therefore introduce the energy-averaged cross
section \cite{Weiler:1982qy,Weiler:1997sh}
\begin{equation}
\langle\sigma_{\nu\bar{\nu}}\rangle\equiv\int\frac{ds}{M_Z^2}
\sigma_{\nu\bar{\nu}}(s)=2\pi\sqrt{2}G_F=40.4\mathrm{nb}\hspace{0.5mm},
\end{equation}
which is the effective cross section for all neutrinos within the
resonance range $(E_{res}(1-\Gamma_Z/M_Z),E_{res}(1+\Gamma_Z/
M_Z))$ and simplify the integration of Eq.(\ref{E:int}) as
\begin{eqnarray}
  &&E_{res}\int_0^\infty\frac{ds}{M_Z^2}F(sE_{res}/M_Z^2)\sigma_{\nu\bar{\nu}}(s)\nonumber\\
  &\simeq&F_0 E_{res}^{1-\alpha}\int_0^\infty\frac{ds}{M_Z^2}\sigma_{\nu\bar{\nu}}(s)\nonumber\\
  &=&F_0 E_{res}^{1-\alpha}\langle\sigma_{\nu\bar{\nu}}\rangle\hspace{0.5mm}.
\end{eqnarray}

Putting everything together we find
\begin{eqnarray}
I_{p|\nu}^Z&=& R_{GZK}n_\nu \xi_{p+n|\nu}
F_0\Big(\frac{M_Z^2}{2m_{\nu}}\Big)^{1-\alpha}\nonumber\\
&\times&\langle\sigma_{\nu\bar{\nu}}\rangle
\mathrm{Br}(Z\rightarrow\mathrm{hadrons})\hspace{0.5mm}.
\end{eqnarray}

The experimental data \cite{PDG} gives the branching ratio
\begin{equation}
\mathrm{Br}(Z\rightarrow\mathrm{hadrons})=(69.89\pm0.07)\%.
\end{equation}
The final proton multiplicity per Z-burst was calculated by Fodor,
Katz and Ringwald(see \cite{FKR:2002hy} and references therein) as
a function of the proton momentum distribution and by Gelmini,
Varieschi and Weiler \cite{GVW:2004zb} using the event generator
PYTHIA \cite{Sjostrand:1993yb}. They obtain
$\xi_{p+n|\nu}\cong2.04$ and $1.6$, respectively. We take the
former value in this paper.

\section{Cascade Limit and Neutrino Mass Bound}
%Cascade upper limit
We now invoke the cascade limit to constrain the UHEC$\nu$ flux
\cite{Astrophysics-of-Cosmic-Rays,cascade-limit}. This is
permissible due to the fact that neutrino productions must always
be accompanied by photons and electrons. The cascades are induced
while these photons or electrons interact with low energy
background radiations such as the CMB in extra galactic space and
the infrared radiation inside the galaxy. The photons so induced
would further cascade and eventually pile up in the energy range
of 10MeV-100GeV with a spectrum $\propto E^{-2}$, which is
consistent with the EGRET observation \cite{Sreekumar:1997un}. The
estimated average energy density in this range is
$\omega_{\mathrm{EGRET}}\approx 2\times10^{-6} \mathrm{eV/cm^3}$.
This provides an upper bound on the UHE neutrino flux,
\begin{equation}
E^2F(E)<\frac{c}{4\pi}\omega_{cas}<\frac{c}{4\pi}\omega_{\mathrm{EGRET}}\hspace{0.5mm}.\label{7}
\end{equation}
To be prudent, we do not assume the exact value of $\alpha=2$ for
the power-law index, but instead leave $\alpha$ as a free
parameter, knowing that its value should be close to 2. Thus the
parameter $F_0$ can be substituted with an upper bound as follows:
\begin{equation}
     \begin{array}{r@{\quad,\quad}l}
      F_0<\frac{c}{4\pi}\omega_{\mathrm{EGRET}}/E_{min}^{2-\alpha} &
      \alpha\geq 2\vspace*{0.5mm}\hspace{0.5mm},\\
      F_0<\frac{c}{4\pi}\omega_{\mathrm{EGRET}}/E_{max}^{2-\alpha} &
      \alpha<2\hspace{0.5mm},\label{EGRET}
     \end{array}
\end{equation}
where $E_{max}$ and $E_{min}$ are the maximum and minimum energies
of the UHEC$\nu$ spectrum.\\

Implementing the cascade limit condition, and inserting all the
relevant physical quantities discussed in the previous section,
Eq.(\ref{main}) becomes
\begin{eqnarray}
  m_\nu>28.7(\frac{E_{res}}{E_{min}})^{\alpha-2}\hspace{0.5mm}\mathrm{eV}
     \hspace{1.5mm},&\hspace{1.5mm}\alpha\geq2, \nonumber\\
  m_\nu>28.7
    (\frac{E_{max}}{E_{res}})^{2-\alpha}\hspace{0.5mm}\mathrm{eV}
     \hspace{1.5mm},&\hspace{1.5mm}\alpha<2\hspace{0.5mm}.\label{neu}
\end{eqnarray}
Note that in this expression not all the $m_{\nu}$ dependence were
grouped to the LHS, as $E_{res}$ clearly depends on $m_{\nu}$.
Nevertheless this expression has an advantage in that
$E_{res}/E_{min}\geq 1$ and $E_{max}/E_{res}\geq 1$ by
definition.\\

An explicit $m_{\nu}$ lower bound can be obtained by moving all
the $m_{\nu}$ dependence to the LHS. We then find
\begin{eqnarray}
m_\nu&>&\frac{1}{2}({\cal A}^{\frac{1}{\alpha-1}}\hspace{0.5mm}
E_{min}^{\frac{2-\alpha}{\alpha-1}}\hspace{0.5mm}M_Z^{2})\hspace{1.5mm},\hspace{1.5mm}\alpha\geq2\hspace{0.5mm},\label{m1}\\
m_\nu&>&\frac{1}{2}({\cal A}^{\frac{1}{\alpha-1}}\hspace{0.5mm}
E_{max}^{\frac{2-\alpha}{\alpha-1}}\hspace{0.5mm}M_Z^{2})\hspace{1.5mm},\hspace{1.5mm}\alpha<2\hspace{0.5mm},\label{m2}
\end{eqnarray}
where
\begin{eqnarray}
\cal{A}&=&I_{>GZK}^{obs}\Big[\frac{c}{4\pi}\omega_{\rm
EGRET}n_\nu R_{GZK} \nonumber\\
&&\xi_{p+n|\nu} \langle\sigma_{\nu\bar{\nu}}\rangle {\rm
Br}(Z\rightarrow{\rm hadrons})\Big]^{-1}.
\end{eqnarray}

The mass bound is dependent on the power-law index $\alpha$ and
the values of $E_{max}$ or $E_{min}$. Our limited knowledge on the
UHEC$\nu$ renders large uncertainty in the determination of
$E_{max}$ and $E_{min}$. One thing which is certain, however, is
that the resonant energy must lie in between $E_{max}$ and
$E_{min}$ for the Z-burst to happen. Eq.(\ref{neu}) indicates that
the minimum value of our bound corresponds to the situation where
either $E_{max}$ or $E_{min}$ equals $E_{res}$, or $\alpha=2$.
Since we should look for the lowest possible lower bound, we put
$E_{res}=E_{max}=E_{min}$ in our estimate and arrive at our
neutrino mass lower bound
\begin{equation}
m_\nu>28.7^{+11.8}_{-10.6}\hspace{0.5mm}{\rm eV}\hspace{5mm}
(R_{max}=R_{GZK}\sim 50{\rm Mpc}),
\end{equation}
where the error comes from fitting the AGASA data
\cite{Nagano:2000ve}.\\

Recent WMAP \cite{Spergel:2003cb} measurement of the CMB
fluctuations has deduced a strong upper limit on neutrino masses,
$\Sigma_i m_{\nu_{ i}}<0.69$eV. Since any single neutrino mass
$\sim 0.04$eV implies a near mass-degeneracy for all three active
neutrinos, one concludes $m_\nu<0.23$eV. Two analyses
 \cite{Barger:2003vs,Allen:2003pt} which include data from WMAP,
2dF, SDSS, and galaxy cluster surveys have arrived at a bound of
$\Sigma_i m_{\nu_{ i}}\lesssim 0.7$eV. Another analysis
 \cite{Crotty:2004gm} using CMB and LSS data gives $\Sigma_i
m_{\nu_{i}}\lesssim 1$eV, but finds a stronger bound $\Sigma_i
m_{\nu_{ i}} \lesssim 0.6$eV when priors from supernova data and
Hubble Key Project are included. These newer results are close to
the original WMAP bound. All these analyses converge to a
cosmological upper bound of $m_\nu\lesssim0.23$eV, which is $2$
order of magnitudes smaller than the lowest possible lower bound
we have derived.

\section{Implication}

Our derivation is based on two assumptions: the saturation of the
observed super-GZK UHECR flux by the Z-burst mechanism and the
cascade upper limit on the maximum UHE$\nu$ flux.\\

Since the cascade limit is deduced from the cascades of photons
accompanying the neutrino production, it is valid for all sources.
Not only astrophysical accelerators (e.g. GRBs, AGNs, SNs, etc.)
but also top-down sources, such as topological defects, superheavy
X particles, dark matter, etc., are all contributing to this limit
as long as the photons are co-produced along-side with neutrinos.
It is generally believed\cite{FKRT:2003ph,Berezinsky:2005rw}  that
the cascade limit on UHE$\nu$ flux is quite robust.\\

Even under the most conservative assumption,
$E_{min}=E_{max}=E_{res}$, i.e. that the UHEC$\nu$ spectrum is a
delta function, our deduced neutrino mass lower bound is more than
120 times larger than the existing upper bound. We therefore
conclude that the Z-burst scenario cannot account for the observed
super-GZK UHECR flux. Assuming all the parameters are fixed, our
neutrino mass lower bound can be lowered if we allow Z-burst to
contribute only partially to the observed UHECR flux. Based on our
values we can conclude that the Z-burst contribution to UHECR
cannot be more than $\sim1\%$ within our local GZK-sphere. The
recent ANITA-lite experiment \cite{ANITA} indicates that the
Z-burst can at best contribute $\sim10\%$ to the UHECR spectrum,
which is consistent with our conclusion.\\

Analogous to the GZK process, Z-burst is one of the few robust
cosmic interaction processes based on the standard model of
particle physics. With a mean-free-path comparable to the present
Hubble radius, it provides much hope to resolve the existing
challenge of the bottom-up scenario. Our negative conclusion on
its viability as a solution to the super-GZK puzzle seems to force
us back to the original dilemma. If the excess super-GZK flux is
found to be real, the need for a solution remains.

%In addition, our study indicates that the neutrino mass upper
%bound can be used to constrain the UHEC$\nu$ flux. For example, a
%less than $\sim1\%$ Z-burst contribution is translated into a
%UHEC$\nu$ flux less than that required for Z-burst to account for
%all super-GZK UHECR flux.

\appendix

\section{A Complete But Ultra-conservative Version}

For completeness, the integration over distance must be carried
out to include the cosmological evolution of the C$\nu$B number
density and the UHEC$\nu$ flux. Since the dependence on the
propagation distance $r$ of the UHEC$\nu$ fluxes and the C$\nu$B
number density can be expressed more straight-forwardly in terms
of the redshift parameter $z$, we make the following change of
variables
\begin{equation}
 dr=-\frac{cH_0^{-1}}{(1+z)\sqrt{\Omega_m(1+z)^3+\Omega_\Lambda}}dz\hspace{0.5mm},
\end{equation}
where $\Omega_m\approx0.3$ and $\Omega_\Lambda\approx0.7$ are the
present matter and dark energy densities in units of the critical
density, respectively.\\

According to the big bang cosmology, the number density of C$\nu$B
is $n_{\nu}(z)=n_\nu(1+z)^3{\rm cm}^{-3}$, where $n_\nu=112{\rm
cm}^{-3}$ is the neutrino-antineutrino number density at present.
The UHEC$\nu$ flux is now assumed to follow a power-law energy
spectrum with cosmological evolution of the source included
\cite{Kalashev:2001sh,Yoshida:1996ie,Yoshida:1998it} and can be
parameterized as
\begin{equation}
F(E,r)=F(E)f(z)=F_0 E^{-\alpha} f_0 (1+z)^{\beta}\hspace{0.5mm},
\end{equation}
where $f_0$ is the normalization factor for the source evolution
determined by the condition
\begin{equation}
\int_0^{z_{max}}f(z)dz=1\hspace{0.5mm},
\end{equation}
with $z_{max}$ being the redshift of the most distant source.\\

A more sophisticated distribution function has been introduced,
based on the star formation rate and the GRB site distribution
\cite{Wick:2003ex}
\begin{equation}
f(z)=f_0\frac{1+a_1}{(1+z)^{-a_2}+a_1(1+z)^{a_3}}\hspace{0.5mm},
\end{equation}
where $a_1, a_2$ and $a_3$ are fitting parameters. It can be shown
that our neutrino mass bound is insensitive between these two
choices of distribution functions and for simplicity we will
invoke Eq.(8) in our subsequent discussion. The separation of
variables allows us to carry out the UHEC$\nu$ energy and distance
integrations independently.\\

%Source evolution
The integration over propagation distance involves several sources
of cosmic evolution,
\begin{eqnarray}
&&\int^{z_{max}}_0 cH_0^{-1}\frac{n_\nu(1+z)^3
 f(z)dz}{(1+z)\sqrt{\Omega_m(1+z)^3+\Omega_\Lambda}}\nonumber\\
&\equiv& cH_0^{-1}n_\nu f(z_{max},\beta)\hspace{0.5mm},
\end{eqnarray}
where simple analytic result is not readily attainable. Table
\ref{evo} displays values of $f(z_{max},\beta)$ for selected
choices of $z_{max}$ and $\beta$. It is clear that
$f(z_{max},\beta)$ is reasonably insensitive to $z_{max}$ and
$\beta$.

\begin{center}
\begin{table}[htb]
 \caption{Values of $f(z_{max},\beta)$ for selected $z_{max}$ and $\beta$.}
 \begin{tabular}{cr|rrrrrrr}\hline\hline
  %\multicolumn{8}{|c|}{\bf cosmological evolution}\\\hline
%\multicolumn{9}{|c|}{$\beta$}\\\hline
 \multicolumn{2}{c}{$\beta$}       & -3 & -2 & -1 & 0 & 1 & 2 & 3 \\\hline
            & 1 & 1.47 & 1.54 & 1.62 & 1.69 & 1.76 & 1.82 & 1.88 \\\cline{2-9}
  $z_{max}$   & 2 & 1.65 & 1.81 & 2.00 & 2.18 & 2.35 & 2.49 & 2.59 \\\cline{2-9}
            & 3 & 1.73 & 1.98 & 2.27 & 2.56 & 2.81 & 2.99 &
            3.12\\\hline\hline
 \end{tabular}\label{evo}
\end{table}
\end{center}

With the knowledge of $f(z_{mzx},\beta)$, the Z-burst yield,
Eq.(\ref{4}), becomes
\begin{eqnarray}
I_{p|\nu}^Z&=& cH_0^{-1}n_\nu f(z_{max},\beta)\xi_{p+n|\nu}
F_0\Big(\frac{M_Z^2}{2m_{\nu}}\Big)^{1-\alpha}\nonumber\\
&\times&\langle\sigma_{\nu\bar{\nu}}\rangle
\mathrm{Br}(Z\rightarrow\mathrm{hadrons})\hspace{0.5mm}.
\end{eqnarray}

Correspondingly, the lower bound on neutrino mass is changed to
\begin{eqnarray}
  m_\nu>1.25f(z_{max},\beta)^{-1}(\frac{E_{res}}{E_{min}})^{\alpha-2}\hspace{0.5mm}\mathrm{eV}
     \hspace{1.5mm},&\hspace{1.5mm}\alpha\geq2, \nonumber\\
  m_\nu>1.25f(z_{max},\beta)^{-1}
    (\frac{E_{max}}{E_{res}})^{2-\alpha}\hspace{0.5mm}\mathrm{eV}
     \hspace{1.5mm},&\hspace{1.5mm}\alpha<2\hspace{0.5mm}.%\label{neu}
\end{eqnarray}

while Eqs.(\ref{m1}) and (\ref{m2}) remain the same form with
\begin{eqnarray}
\cal{A}&=&I_{>GZK}^{obs}\Big[\frac{c}{4\pi}\omega_{\rm
EGRET}cH_0^{-1}n_\nu \nonumber\\
&&\xi_{p+n|\nu} \langle\sigma_{\nu\bar{\nu}}\rangle {\rm
Br}(Z\rightarrow{\rm hadrons})\Big]^{-1}.
\end{eqnarray}

The mass bound is now dependent on the numerical values of the
evolution factor $f(z_{max},\beta)$ as well (See Table \ref{t2}).
Again we look for the lowest possible lower bound by putting
$E_{res}=E_{max}=E_{min}$ in our estimate. As we have shown,
$f(z_{max},\beta)$ is of the order $1$ and is insensitive to
$z_{max}$ and $\beta$. With the choice of
$f(z_{max},\beta)=f(3,0)=2.56$, we arrive at our neutrino mass
lower bound
\begin{equation}
m_\nu>0.49^{+0.20+0.53}_{-0.18-0.14}\hspace{0.5mm}{\rm
eV}\hspace{10mm} (R_{max}\sim cH_0^{-1}).
\end{equation}
The former error comes again from the AGASA data and the latter
from the uncertainty of the evolution factor $f(z_{max},\beta)$.\\

%Table of mass bound
\begin{center}
\begin{table}[htb]
\caption{Values of mass lower bound for selected $z_{max}$ and
$\beta$ with $\alpha=2$ energy spectrum.}
\begin{tabular}{crrrrrrrrr}\\\hline\hline
\multicolumn{10}{c}{$\alpha=2$}\\\hline\hline
  $\beta$&-3&-3&-3& 0& 0& 0& 3& 3& 3\\
  $z_{max}$         &   1&   2&   3&   1&   2&   3&   1&   2&   3\\
  $f(z_{max},\beta)$&1.47&1.65&1.73&1.69&2.18&2.56&1.88&2.59&3.12\\
  mass
  bound             &0.85&0.76&0.73&0.74&0.58&0.49&0.68&0.48&0.40\\\hline\hline
\end{tabular}\label{t2}
\end{table}
\end{center}

\acknowledgements

We thank P. Blasi, Je-An Gu and K. Reil for valuable discussions.
KCL appreciates guidance and support of W-Y.\ P.\ Hwang. This work
is supported by the National Science Council (NSC
94-2112-M-002-029; NSC 93-2112-M-002-025) of Taiwan, R.O.C., and
in part by US Department of Energy under Contract No.\
DE-AC03-76SF00515.

%\bibliographystyle{unsrt}
%\bibliography{EHENeu}

\end{document}